\documentclass[11pt]{article}

\usepackage[utf8]{inputenc}
\usepackage[T1]{fontenc}
\usepackage{amsmath,amssymb}
\usepackage{graphicx}
\usepackage{booktabs}
\usepackage{tabularx}
\usepackage{float}
\usepackage{subcaption}
\usepackage{changepage}      
\usepackage[colorlinks=true,citecolor=blue,linkcolor=blue,urlcolor=blue]{hyperref}
\usepackage[numbers,sort&compress]{natbib}
\usepackage{xspace}
\usepackage[margin=1in]{geometry}
\usepackage{orcidlink}
\usepackage{authblk}

\newcolumntype{C}{>{\centering\arraybackslash}X}

\newcommand{\rewsym}[1]{\ensuremath{\mathcal{R}_{\text{#1}}}}

\newcommand{\IGM}{\textsc{IGM}\xspace}
\newcommand{\CTDE}{\textsc{CTDE}\xspace}
\newcommand{\QMIX}{\textsc{QMIX}\xspace}
\newcommand{\IQL}{\textsc{IQL}\xspace}

\title{Cooperative EV Charging Assignment Using Multi-Agent Reinforcement Learning: A Comparison of QMIX and Independent Q-Learning Under Urban Traffic Uncertainty}

\author[1]{Nour-Eddine Moumni\orcidlink{0009-0008-5803-0604}\thanks{Corresponding author: \texttt{noureddine\_moumni@um5.ac.ma}}}
\author[1,2]{Rachid Alaoui\orcidlink{0000-0001-9948-329X}}
\author[3]{Driss Kiouach\orcidlink{0000-0002-1497-3028}}

\affil[1]{Laboratory of Research in Computer Science and Telecommunications (LRIT), Mohammed V University, Rabat, Morocco}
\affil[2]{Higher School of Technology of Sal\'e (EST Sal\'e), Mohammed V University, Sal\'e, Morocco; \texttt{rachid.alaoui@est.um5.ac.ma}}
\affil[3]{L2MASI Laboratory, Faculty of Sciences Dhar El Mahraz, Sidi Mohamed Ben Abdellah University, Fez, Morocco; \texttt{kiouach@gmail.com}}

\date{}

\begin{document}
\maketitle

\begin{center}
\emph{Presented at ICATH 2026 (Rabat, Morocco, 15--17 July 2026); accepted for MDPI Engineering Proceedings (in press).}
\end{center}

\begin{abstract}
The allocation of electric vehicles to public charging stations is increasingly constrained by the fact that several heterogeneous facilities often serve the same stream of requests while observing only partial local information. In such a setting, an accept/reject decision that is defensible for one station may nevertheless reduce the efficiency of the charging district as a whole. The present study formulates this operational problem as a Decentralized Partially Observable Markov Decision Process (Dec-POMDP) and examines whether Centralized Training with Decentralized Execution (CTDE), instantiated through QMIX, can improve assignment performance relative to a cooperative Independent \textsc{Q-Learning} baseline (IQL). The evaluation is conducted in a six-station urban network simulated with SUMO, using five independent random seeds, 200 training episodes, and four traffic flux scenarios ranging from 100 to 250 vehicles/hour. Under this protocol, QMIX reaches a mean reward per evaluated decision of $+0.776$, compared with $+0.578$ for IQL ($\Delta = +0.198$, seed-paired Cohen's $d_z = 2.98$; exact two-sided Wilcoxon signed-rank test across the five seeds, $p = 0.0625$, the minimum attainable value at this sample size), and increases the acceptance rate over all evaluated requests from 72.4\% to 94.0\% ($\Delta = +21.6$~pp, $d_z = 5.24$, Wilcoxon $p = 0.0625$). A five-variant ablation suggests this gain is primarily attributable to the CTDE architecture when holding all cooperative reward terms fixed, though not formally isolated with a single seed (a coincidentally similar single-seed $\Delta = +0.198$, $d = 2.46$, reported descriptively since the 50 episodes come from one seed). By contrast, cooperative reward shaping ($L_{\text{coop}} = 0.20$) and experience sharing ($R_{\text{exp}} = 0.15$) show no clear descriptive benefit and may slightly reduce performance. These findings support the view that, for multi-station charging assignment, coordination should be embedded in the learning architecture rather than added only through heuristic cooperation terms.
\end{abstract}


\section{Introduction}
\label{sec:intro}

In urban electric-mobility systems, assigning vehicles to public charging stations is a problem of distributed coordination under incomplete information, not merely capacity allocation. A vehicle may be associated with several candidate stations, each with a different charging rate, queue state, and occupancy level, yet the accept/reject decision is made locally: no station observes the complete network state, so system-level quality of service emerges from decentralized decisions coupled through congestion, waiting time, and availability. Reinforcement learning has been used in charging control for over a decade~\citep{ref-vandael2015}, but early work targeted single-station or fleet-level formulations where standard methods suffice~\citep{ref-cao2021,ref-vandael2015}. Extending to several stations motivates Independent \textsc{Q-Learning} (\IQL) as a natural baseline~\citep{ref-tan1993}, while exposing the limits of treating each station as an isolated learning problem.

Under independent learning, each station estimates its own Q-function while the other stations' policies keep changing, so the environment is non-stationary from any single learner's perspective, weakening the stationarity assumption behind tabular \textsc{Q-Learning}. The resulting policies can be locally defensible yet network-inefficient, since one station's accept/reject decision shifts the demand and queue pressure faced by the others. This coordination loss is reflected here in a $21.6$\,pp acceptance-rate gap, from $72.4\%$ (\IQL{}) to $94.0\%$ (\QMIX{}).

Centralized Training with Decentralized Execution (\CTDE{}) addresses this mismatch between local observability and global performance, within the Dec-POMDP framework~\citep{ref-oliehoek}: joint state information stabilizes training, while decentralized execution and the absence of inter-station messaging are preserved at deployment. \QMIX{}~\citep{ref-rashid2018} implements this via a monotonic mixing network that combines individual action-value estimates into a joint value function, the monotonicity constraint preserving the Individual-Global-Max (\IGM) property so that the centralized-value-maximizing action can still be recovered through decentralized per-agent maximization.

\begin{figure}[!htb]
\centering
\includegraphics[width=\linewidth]{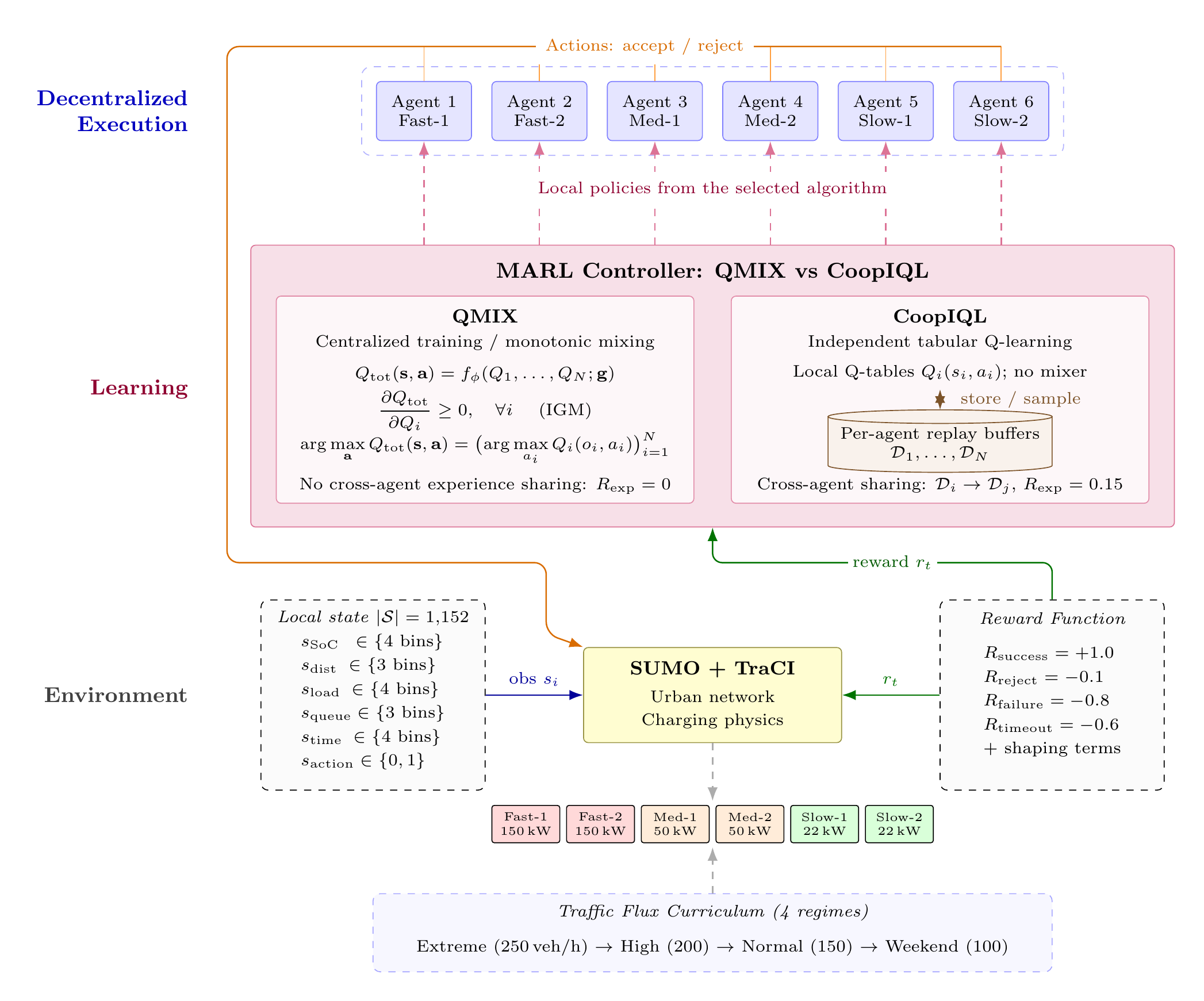}
\caption{Overview of the cooperative MARL framework. A centralized MARL controller trains six station agents (two fast, two medium, two slow) on a SUMO-simulated urban network through \CTDE{}. Each agent observes a local $1{,}152$-state discretization and selects accept/reject in real time. Training cycles through four traffic flux regimes; rewards combine success/failure outcomes with shaping terms. The figure contrasts the centralized training phase (purple) with the decentralized execution phase (blue).\label{fig:architecture}}
\end{figure}

Figure~\ref{fig:architecture} summarizes the framework: a centralized controller trains six station agents (two fast, two medium, two slow) under \CTDE{} on a SUMO-simulated urban network, each observing a local 1,152-state discretization and selecting accept/reject in real time across four traffic flux regimes.

Although \CTDE{} has been adopted in related EV energy-planning studies~\citep{ref-mishra2025,ref-wang2025}, several aspects of its use for charging station allocation remain insufficiently characterized: the magnitude of the \CTDE{}-vs-independent-learning gap is rarely estimated under a multi-seed protocol, the origin of any cooperative gain is often unidentifiable because architecture, reward shaping, experience sharing, and implementation choices are confounded, and whether the advantage holds across traffic regimes and station types remains untested.

To address these issues, we adopt a controlled empirical protocol: \QMIX{} and \IQL{} are compared on the same six-station task over five seeds and 200 episodes, with results reported via statistical tests and effect sizes; a five-variant ablation isolates the effects of CTDE, cooperative reward shaping, and experience sharing; and the analysis is stratified by traffic regime and station type to assess the stability of the observed advantage.

\section{Related Work}
\label{sec:related}

\subsection{Reinforcement Learning for EV Charging Management}

Research on RL for EV charging has moved from simple decision settings toward increasingly distributed formulations. Early contributions considered single-station or fleet-level problems, showing that tabular or batch RL could learn day-ahead charging policies without exact flexibility models~\citep{ref-vandael2015}, and subsequent online actor--critic approaches incorporated arrival, departure, and demand uncertainty into real-time control~\citep{ref-cao2021}. In parallel, the MARL literature has established that multiple interacting decision makers change the nature of the learning problem itself: non-stationarity, partial observability, scalability, and credit assignment become structural issues rather than peripheral implementation difficulties~\citep{ref-hernandez2019}. Distributed charging assignment across heterogeneous stations belongs to this class of problems.

Two bodies of work are worth distinguishing here: methodological contributions to MARL algorithms themselves, and the use of these algorithms in EV charging and energy-management contexts. The distinction matters because an observed improvement in an application study may originate from the learning architecture, the physical modeling assumptions, the reward design, or constraints imposed by the simulator.

Recent EV-oriented studies have accordingly moved toward distributed or multi-station formulations. Korkas et al.~\citep{ref-korkas2024} proposed a distributed MARL framework coordinating multiple charging spots within a station; Jiang et al.~\citep{ref-jiang2025} studied profit maximization across multiple charging stations using QMIX-based coordination; Fan et al.~\citep{ref-wang2025} examined decentralized energy management of chargers within a solar-assisted charging station under uncertainty. Closer to the present assignment setting, Suanpang and Jamjuntr~\citep{ref-suanpang2024} compare DDPG, DQN, and MADDPG for EV charging-station recommendation in smart cities, finding that MADDPG's centralized-critic/decentralized-actor design performs best; the present study differs by using a comparable CTDE-vs-independent-learning contrast as a controlled \emph{attribution} study -- isolating, via ablation, whether the gain stems from the CTDE architecture itself rather than from auxiliary cooperative mechanisms -- rather than ranking algorithms for a recommendation task. These studies show multi-agent learning can improve coordination and cost outcomes, but remain focused on grid-side or operator-side objectives such as profitability, energy cost, and station-level scheduling.

Complementary work has explored infrastructure monitoring and predictive control through IoT-based sensing and load-forecasting~\citep{ref-martins2025,ref-mohsenimanesh2024}, improving observability but relying on deterministic scheduling or forecasting modules rather than online multi-station assignment under stochastic urban traffic. User-facing charging assignment thus remains less systematically examined, particularly the question of which cooperative mechanism, not merely whether cooperation, drives the observed gains.

\subsection{CTDE Approaches in EV Charging}

CTDE was introduced to satisfy two constraints that are difficult to reconcile in cooperative MARL simultaneously: the need for joint information during learning and the requirement that each agent execute locally. Centralized information is used only while values or policies are fitted; once training is complete, the station-level decision rule must remain implementable from local observations alone. VDN~\citep{ref-sunehag2018} represents the joint value as an additive composition of individual values; QMIX~\citep{ref-rashid2018} relaxes this by learning a monotonic mixing function, representing richer joint effects while preserving the Individual Global Maximum (IGM) condition, compatible with the Dec-POMDP view of decentralized decision making under imperfect information~\citep{ref-oliehoek}.

In EV applications, CTDE has mainly appeared where the objective is defined from the operator or energy-system perspective: anticipatory charging with BiLSTM-based forecasting~\citep{ref-mishra2025}, profit-oriented coordination across charging regions~\citep{ref-jiang2025}, and decentralized control of stations coupled with photovoltaic generation~\citep{ref-wang2025}. These studies show joint training is useful in charging and energy-management systems, but do not answer the assignment question considered here, where vehicles arrive from an urban traffic process and station decisions directly determine acceptance.

A further difficulty is that several design choices are often introduced together, so a reported gain may stem from CTDE, an engineered reward term, a forecasting block, or simulator assumptions, making joint credit assignment hard to isolate --- particularly in urban charging networks, where demand fluctuations, station heterogeneity, and partial observability act simultaneously. The present study therefore treats CTDE as a component whose contribution must be separated from reward shaping and experience sharing, not merely as a performance mechanism.

\subsection{Positioning of the Present Work}

The study is accordingly designed as a controlled \emph{attribution study} rather than an algorithm-design contribution, evaluated on a heterogeneous six-station urban network under a common protocol and metric set. Relative to the MARL literature, the work sits on the application and empirical-evaluation side; relative to the EV charging literature, it provides a decomposition usually left implicit --- separating cooperative performance into architectural and heuristic components. The relevant question is therefore not whether cooperation helps in the abstract, but which element of the cooperative system drives the measured improvement. The results suggest a QMIX advantage under the studied protocol, and the single-seed ablation provides a descriptive indication that CTDE is the component responsible, while cooperative reward shaping and experience sharing are neutral or detrimental. To our knowledge, this component-level attribution has not previously been established with comparable statistical discipline in the EV charging assignment literature.

\section{System Architecture and Methods}
\label{sec:methods}

\subsection{Problem Formulation}

We model EV charging assignment as a Decentralized Partially Observable Markov Decision Process (Dec-POMDP)~\citep{ref-oliehoek} with $N = 6$ station agents, where at each decision step, when vehicle $v$ requests charging, station $i$ observes local state $\mathbf{s}_i \in \mathcal{S}$ and selects action $a_i \in \{\texttt{accept}, \texttt{reject}\}$, the joint action $\mathbf{a} = (a_1, \ldots, a_6)$ determines which station --- if any --- serves the vehicle while stations that reject receive no service commitment, and a centralized reward shaper computes $r_t$ from the outcome (success, rejection, failure, timeout) and broadcasts it to all active agents.

The local observation $\mathbf{s}_i$ encodes six discrete features:

\begin{equation}
\mathbf{s}_i = \bigl(\underbrace{s_{\text{SoC}}}_{\times 4},\;
\underbrace{s_{\text{dist}}}_{\times 3},\;
\underbrace{s_{\text{load}}}_{\times 4},\;
\underbrace{s_{\text{queue}}}_{\times 3},\;
\underbrace{s_{\text{time}}}_{\times 4},\;
\underbrace{s_{\text{action}}}_{\times 2}\bigr),
\label{eq:state}
\end{equation}

\noindent where $s_{\text{SoC}} \in [0\text{--}20\%, 20\text{--}40\%, 40\text{--}60\%, 60\text{--}100\%]$, $s_{\text{dist}} \in [0\text{--}500\text{m}, 500\text{--}1000\text{m}, {>}1000\text{m}]$, $s_{\text{load}} \in \{0,1,2,3+\}$, $s_{\text{queue}} \in \{0, 1\text{--}3, {\geq}4\}$, $s_{\text{time}} \in [0\text{--}15\text{min}, 15\text{--}30\text{min}, 30\text{--}45\text{min}, {>}45\text{min}]$, and $s_{\text{action}} \in \{0,1\}$ encodes the previous routing decision, yielding a discrete state space of $|\mathcal{S}| = 1{,}152$ states (empirically confirmed via model-file metadata across all seeds).

\subsection{Reward Function}

The reward function is shared across IQL and QMIX to ensure strict comparability, with base outcomes $\rewsym{success} = +1.0$ (vehicle successfully initiates charging), $\rewsym{reject} = -0.1$ (controlled rejection), $\rewsym{failure} = -0.8$ (accepted but service fails), and $\rewsym{timeout} = -0.6$ (vehicle does not arrive in time), three shaping terms penalizing queue buildup ($L_q = 0.20$), excessive routing distance ($L_d = 0.10$), and station saturation ($L_s = 0.30$), and one term rewarding urgently-low-SoC vehicles ($L_{\text{SoC}} = 0.10$); for cooperative variants A0, A2, and A3, a load-balance bonus

\begin{equation}
r_{\text{coop},i} = L_{\text{coop}}
    \cdot \frac{\bar{\ell} - \ell_i}{C},
\label{eq:coop}
\end{equation}

\noindent rewards station $i$ for accepting when it carries below-average load ($\bar{\ell}$ = fleet mean load, $\ell_i$ = station $i$'s current load, $C$ = capacity), bounded in $[-L_{\text{coop}}, +L_{\text{coop}}]$, with $L_{\text{coop}} = 0.20$.

\subsection{Algorithm Descriptions}

\textbf{IQL (Cooperative Independent \textsc{Q-Learning}).}
IQL follows the independent-learner setting introduced by Tan~\citep{ref-tan1993}: each of the $N = 6$ station agents maintains its own tabular $Q_i(s_i,a_i)$, updated via the standard Q-learning rule of Watkins and Dayan~\citep{ref-watkins1992q}, with $\alpha = 0.15$, $\gamma = 0.95$, and $\varepsilon_{\min} = 0.08$.

Two mechanisms make this baseline cooperative rather than purely isolated: experience sharing ($R_{\text{exp}} = 0.15$), inserting a fraction of each agent's transitions into other agents' replay buffers, and the cooperative bonus of Equation~(\ref{eq:coop}), encouraging network-level load balancing. These give QMIX a natural comparison point representing cooperation through heuristic exchange and reward shaping rather than centralized value factorization; as the ablation shows, however, both interact unfavorably with independent updates in the non-stationary multi-station environment.

\textbf{QMIX (Proposed).}
QMIX~\citep{ref-rashid2018} replaces the tabular independent learners with a deep value-factorization architecture implementing \CTDE{}: global state is available to the mixing network during training, while execution uses only local observations. Each agent network estimates $Q_i(o_i,a_i)$, and the centralized mixer combines these into a joint $Q_{\text{tot}}(\mathbf{s},\mathbf{a})$.

The mixing network is constrained to be monotonic with respect to the individual utilities,
\[
\frac{\partial Q_{\text{tot}}}{\partial Q_i} \geq 0,\quad \forall i,
\]
which preserves the Individual-Global-Max (IGM) condition:
\[
\arg\max_{\mathbf{a}} Q_{\text{tot}}(\mathbf{s},\mathbf{a})
=
\bigl(
\arg\max_{a_1} Q_1(o_1,a_1),\dots,
\arg\max_{a_N} Q_N(o_N,a_N)
\bigr).
\]
This property is essential for decentralized execution because the joint greedy action can be recovered from the greedy actions of the individual station policies.

The mixing network's weights are generated by hypernetworks~\citep{ref-ha2016hypernetworks} conditioned on the global state vector $\mathbf{g} \in \mathbb{R}^{25}$, which concatenates the six agents' occupancy, queue, and load features together with the current flux regime and last joint action --- letting the value factorization depend on network-level information while preserving decentralized action selection at deployment. The six agent networks share weights (hidden dimension $d_h = 64$), reducing the agent-network parameterization by a factor of $6$ relative to fully separate networks.

Training uses the Adam optimizer~\citep{ref-kingma2015adam} (learning rate $5 \times 10^{-4}$, batch size $32$, $\gamma = 0.99$), with soft target updates ($\tau = 0.005$). Unlike IQL, explicit experience sharing is disabled: the centralized mixer already provides joint credit assignment through the global training signal, and injecting off-policy transitions from heterogeneous station trajectories would add mismatch rather than reliable coordination.

\subsection{Simulation Environment}

The SUMO microscopic traffic simulator~\citep{ref-lopez2018} is interfaced via TraCI to provide realistic vehicle dynamics, route-level travel times, and station-level charging physics across an urban network of six heterogeneous stations --- two fast ($50\,\text{kW}$), two medium ($22\,\text{kW}$), and two slow ($11\,\text{kW}$) --- each with a hard capacity of three concurrent vehicles, while four traffic curricula constitute the training regime: \emph{Extreme} (250\,veh/h), \emph{High} (200\,veh/h), \emph{Normal} (150\,veh/h), and \emph{Weekend} (100\,veh/h), applied sequentially for the first 25 episodes and cycled randomly thereafter, with charging durations scaling proportionally with vehicle energy needs and station power and on-the-fly SoC derating above $80\%$ to reflect real charger behavior.

\subsection{Experimental Protocol and Ablation Design}

We evaluate both algorithms over 200 training episodes for each of 5 independent seeds (42, 99, 777, 1234, 2024), which results in 261,467 vehicle-level decision records overall. This multi-seed setting is used for the main comparison reported in the next section.

The ablation protocol is deliberately restricted relative to the main benchmark. A single seed (42) is used and each variant is run for 50 episodes, following the design summarized in Table~\ref{tab:ablation}. The purpose of this reduced protocol is not to reproduce the full performance estimate, but to compare the effect of three controlled components under identical task conditions: CTDE, cooperative reward shaping, and experience sharing.

\begin{table}[!htb]
\caption{Ablation design used to isolate the contribution of CTDE, cooperative reward shaping, and experience sharing. Each variant differs from the reference configuration by at most two controlled changes.\label{tab:ablation}}
\centering
\begin{tabularx}{\textwidth}{lllCCCr}
\toprule
\textbf{ID} & \textbf{Variant} & \textbf{Algo} &
\textbf{Coop} & $L_{\text{coop}}$ &
$R_{\text{exp}}$ & \textbf{Reward}\\
\midrule
A0 & QMIX (full)           & QMIX & \checkmark & 0.20 & 0.00 & $+0.759$ \\
A1 & QMIX (without coop)   & QMIX & $\times$   & 0.00 & 0.00 & $+0.809$ \\
\midrule
A2 & IQL (baseline)        & IQL     & \checkmark & 0.20 & 0.15 & $+0.517$ \\
A3 & IQL (without sharing) & IQL     & \checkmark & 0.20 & 0.00 & $+0.561$ \\
A4 & IQL (pure)            & IQL     & $\times$   & 0.00 & 0.00 & $+0.584$ \\
\bottomrule
\end{tabularx}
\noindent{\footnotesize{%
Pairwise comparisons used for attribution: A0 vs.\ A1 (effect of cooperative reward shaping in QMIX); A2 vs.\ A3 (effect of experience sharing in IQL); A3 vs.\ A4 (effect of cooperative reward shaping alone in IQL); A0 vs.\ A3 (effect of CTDE, holding $L_{\text{coop}}$ and $R_{\text{exp}}$ fixed).}}
\end{table}

\vspace{0.3em}
This design constitutes a compact attribution study in which each comparison modifies one element while the remaining experimental conditions are held fixed. Because the 50 episodes of each variant come from a single seed rather than independent replicates, the ablation contrasts are reported descriptively, as reward differences ($\Delta$) and effect sizes (Cohen's $d$), without a significance test.

\section{Results}
\label{sec:results}

\subsection{Main Performance Comparison}
\label{subsec:main}

Table~\ref{tab:main} reports the quality-of-service metrics, pooled over every evaluated decision across the five seeds and 200 training episodes. Under this protocol, QMIX achieves a mean reward per decision of $+0.776$, whereas IQL reaches $+0.578$ ($\Delta = +0.198$, seed-paired Cohen's $d_z = 2.98$; exact Wilcoxon signed-rank $p = 0.0625$, $n = 5$ seeds). The same pattern is observed for the acceptance rate, where QMIX reaches $94.0\%$ compared with $72.4\%$ for IQL ($\Delta = +21.6\,\text{pp}$, $d_z = 5.24$, Wilcoxon $p = 0.0625$).

\begin{table}[!htb]
\caption{QoS comparison, pooled over every evaluated decision across 5 seeds $\times$ 200 episodes. Best value per metric marked with $\dagger$; ns = not significant.\label{tab:main}}
\centering
\begin{tabularx}{\textwidth}{lCCCl}
\toprule
\textbf{Metric} & \textbf{IQL} & \textbf{QMIX} &
$\Delta$ & \textbf{Stat.}\\
\midrule
Mean Reward             & $+0.578$ & $+0.776\,\dagger$ &
    $+0.198$ & $d_z = 2.98$$^{\ddagger}$ \\
Accept Rate (\%)        & $72.4$   & $94.0\,\dagger$   &
    $+21.6\,\text{pp}$ & $d_z = 5.24$$^{\ddagger}$ \\
Success/Accept (\%)     & $100.0$  & $100.0$           & $0.0$   & --- \\
Travel Time (s)         & $13.3\,\dagger$   & $15.9$   &
    $+2.5$ & selection bias $^{*}$ \\
Waiting Time (s)        & $51.4$   & $47.9\,\dagger$   & $-3.5$ & ---\\
Load Balance CV         & $0.531$  & $0.551$           & $+0.020$ & ns \\
\bottomrule
\end{tabularx}
\noindent{\footnotesize{%
$^{*}$ QMIX accepts more distant vehicles, so the travel time increase is expected; both algorithms maintain 100\% success-to-accept, confirming no speculative over-acceptance. $^{\ddagger}$ Effect size ($d_z$) from a paired comparison across the $n = 5$ independent seeds (the true number of replicates). The reported $p$-value is the exact two-sided Wilcoxon signed-rank test, the smallest attainable at this sample size; a parametric paired $t$-test gives $p = 0.003$ (reward) and $p < 0.001$ (acceptance) but assumes near-normality of only five paired differences.}}
\end{table}

\begin{figure}[!htb]
\centering
\includegraphics[width=\linewidth]{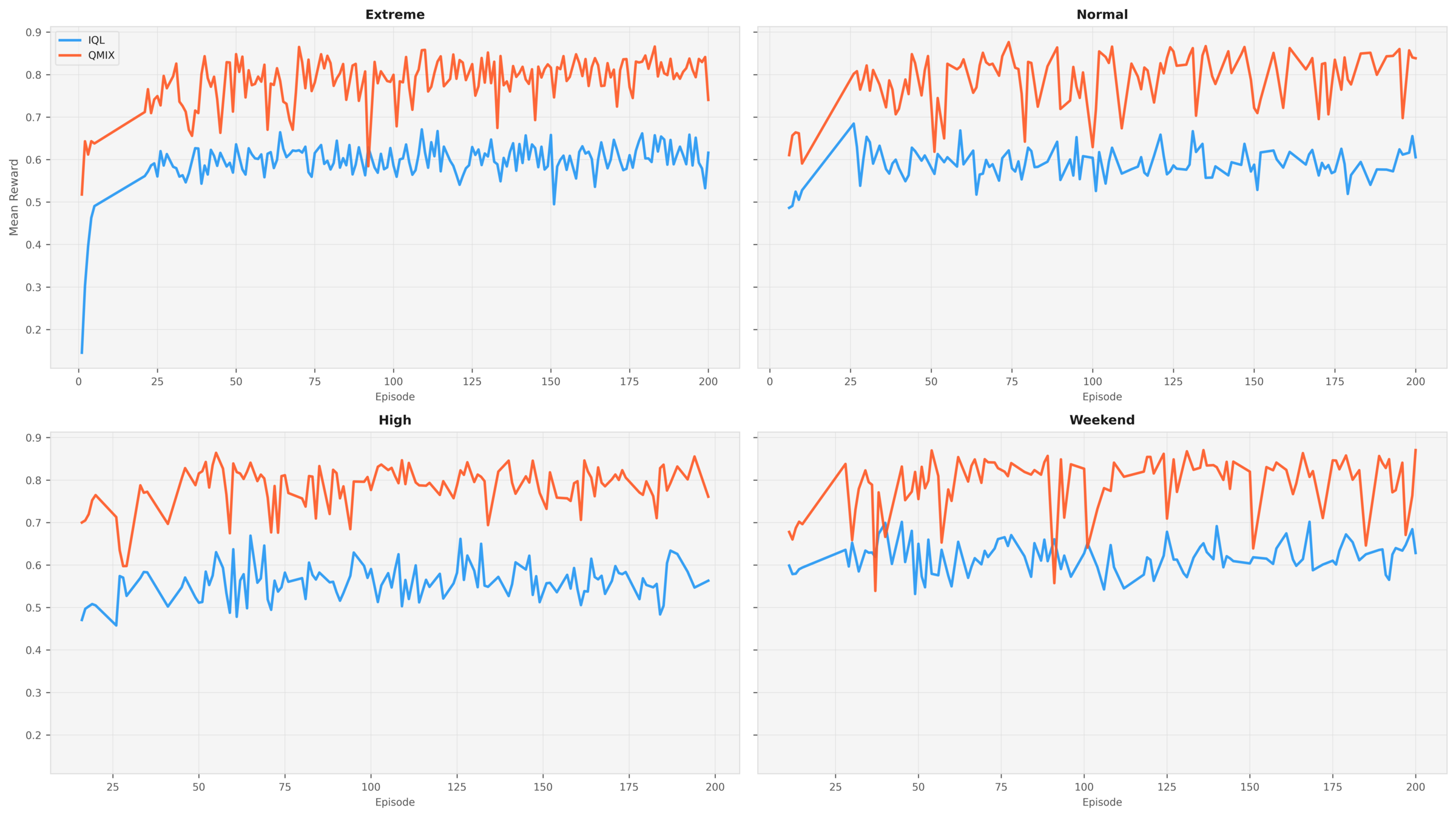}
\caption{Reward convergence curves (IQL = blue, QMIX = orange) per traffic flux scenario, showing per-episode mean reward across 5 seeds. QMIX converges faster and achieves higher asymptotic performance in all scenarios, with the largest absolute gap under \emph{High} flux where station contention is greatest.\label{fig:convergence}}
\end{figure}

The convergence profiles of Figure~\ref{fig:convergence} add the temporal dimension: QMIX reaches near-plateau performance within the first 25~episodes across all four flux scenarios, whereas IQL is slower and more volatile, reflecting non-stationary updates across six simultaneously learning agents, with the largest gap under \emph{High} flux (200\,veh/h), where inter-station contention is greatest (Section~\ref{sec:discussion}).

\subsection{Performance Across Traffic Flux Scenarios}
\label{subsec:flux}

\begin{figure}[!htb]
\centering
\includegraphics[width=\linewidth]{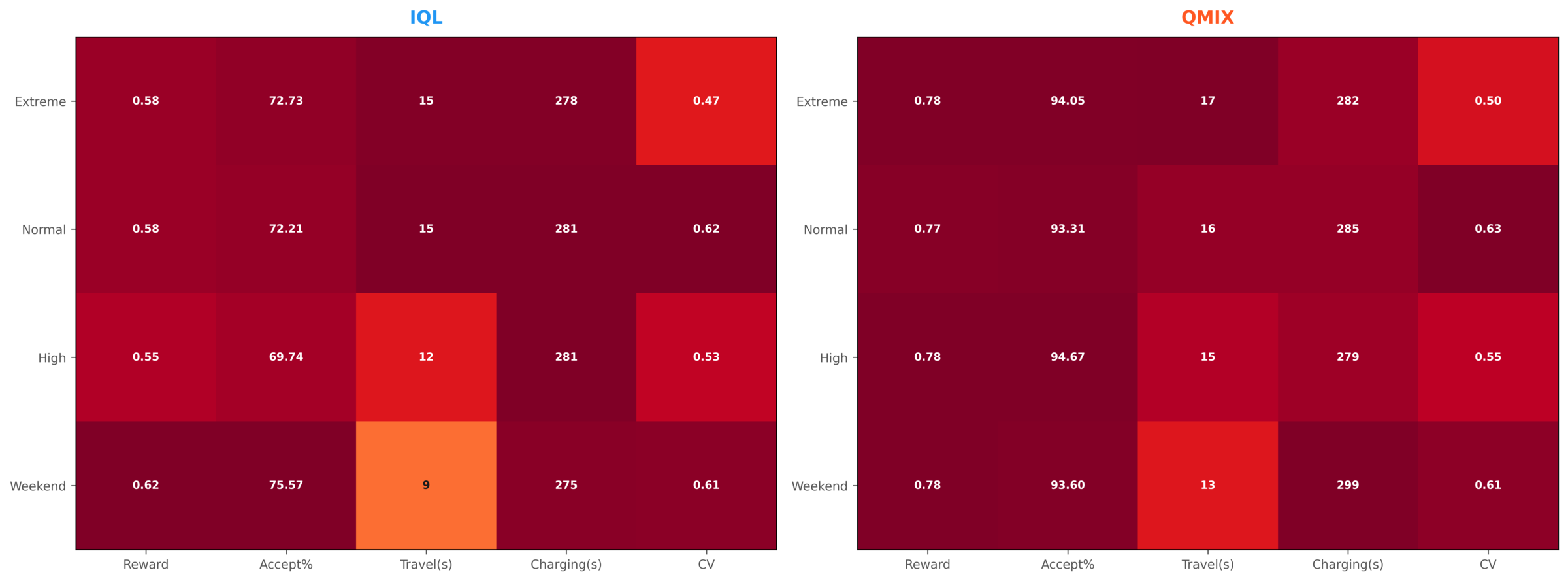}
\caption{Performance heatmap across four traffic flux scenarios (rows) and five QoS dimensions (columns) for IQL (left) and QMIX (right), where cell values are raw means and color intensity is column-normalized. QMIX dominates on reward and accept rate across all scenarios, while load balance CV remains comparable.\label{fig:heatmap}}
\end{figure}

Figure~\ref{fig:heatmap} disaggregates performance by flux scenario across five QoS dimensions: reward advantage is consistent across all regimes, from $\Delta = +0.165$ (Weekend) to $\Delta = +0.230$ (High), with the accept-rate gap peaking at $+25\,\text{pp}$ under \emph{High} flux. Two secondary patterns emerge: travel time is consistently higher for QMIX by roughly 1--5\,s across regimes (smallest under \emph{Normal} flux, largest under \emph{Weekend}), a structural consequence of its broader acceptance radius rather than routing inefficiency, and load balance CV is comparable between algorithms across all regimes (clarified by the ablation below).

\subsection{Station-Level Analysis}
\label{subsec:station}

\begin{figure}[!htb]
\centering
\includegraphics[width=\linewidth]{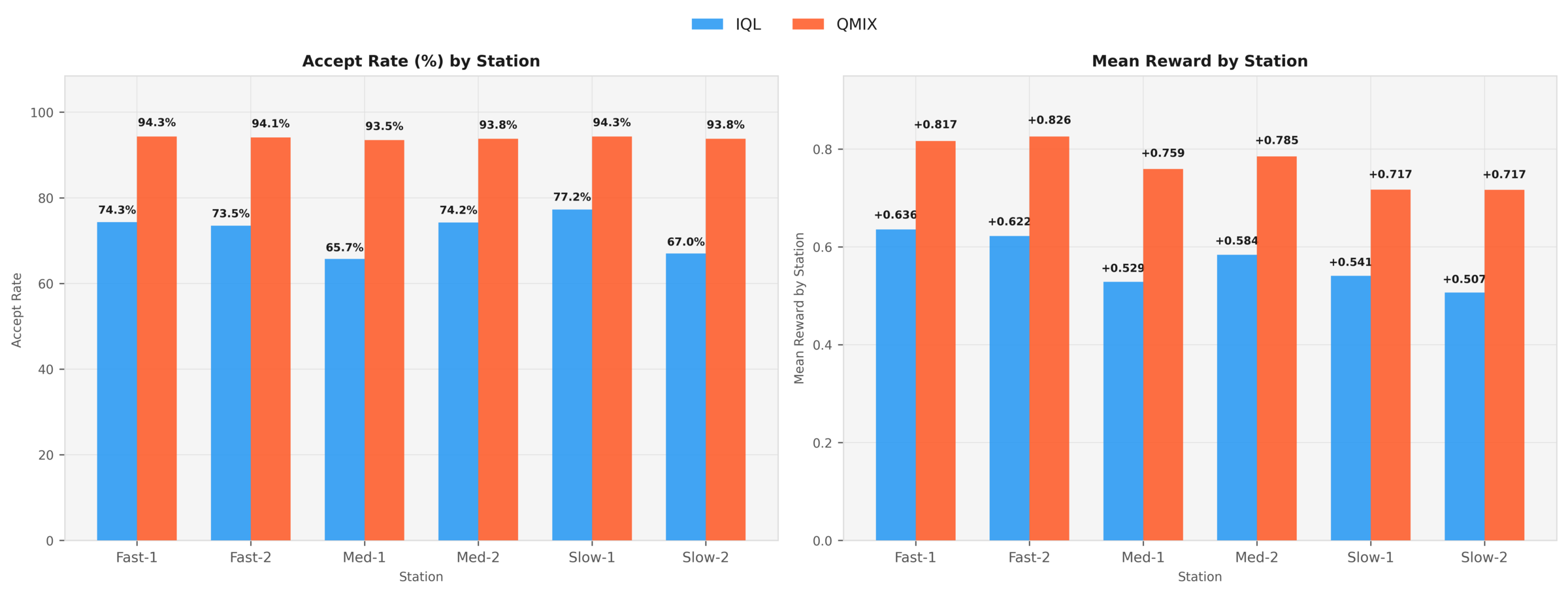}
\caption{Accept rate (left) and mean reward (right) by station type for IQL and QMIX (5 seeds $\times$ 200 episodes), where QMIX achieves uniformly high accept rates across all station types while IQL performance is most degraded at Med-1 and Slow-2, the two stations with the lowest accept rate under contention.\label{fig:stations}}
\end{figure}

Figure~\ref{fig:stations} shows IQL's performance is not uniformly poor but \emph{unevenly} poor: accept rates range from $65.7\%$ at Med-1 to $77.2\%$ at Slow-1 (spread of $11.5\,\text{pp}$), reflecting differential exposure to contention across station types. QMIX instead achieves $93.5$--$94.7\%$ uniformly (spread of just $1.2\,\text{pp}$), as the mixer learns a globally-consistent acceptance policy that neutralizes the asymmetry baked into heterogeneous station capacities --- a property independent learners cannot achieve without explicit coordination.

\subsection{Ablation Study}
\label{subsec:ablation}

\begin{figure}[!htb]
\centering
\begin{subfigure}[t]{0.48\linewidth}
\centering
\includegraphics[width=\linewidth]{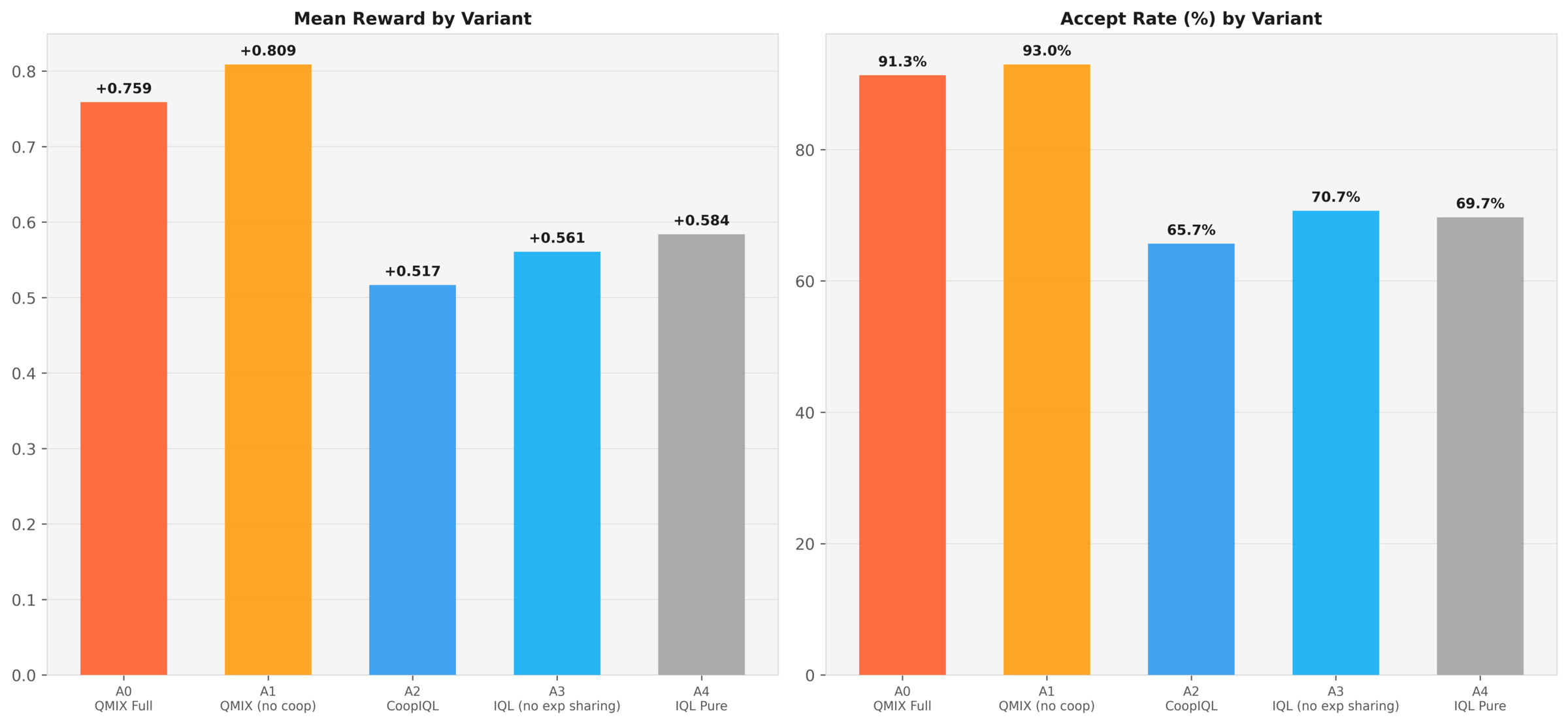}
\caption{Variant comparison}
\end{subfigure}
\hfill
\begin{subfigure}[t]{0.48\linewidth}
\centering
\includegraphics[width=\linewidth]{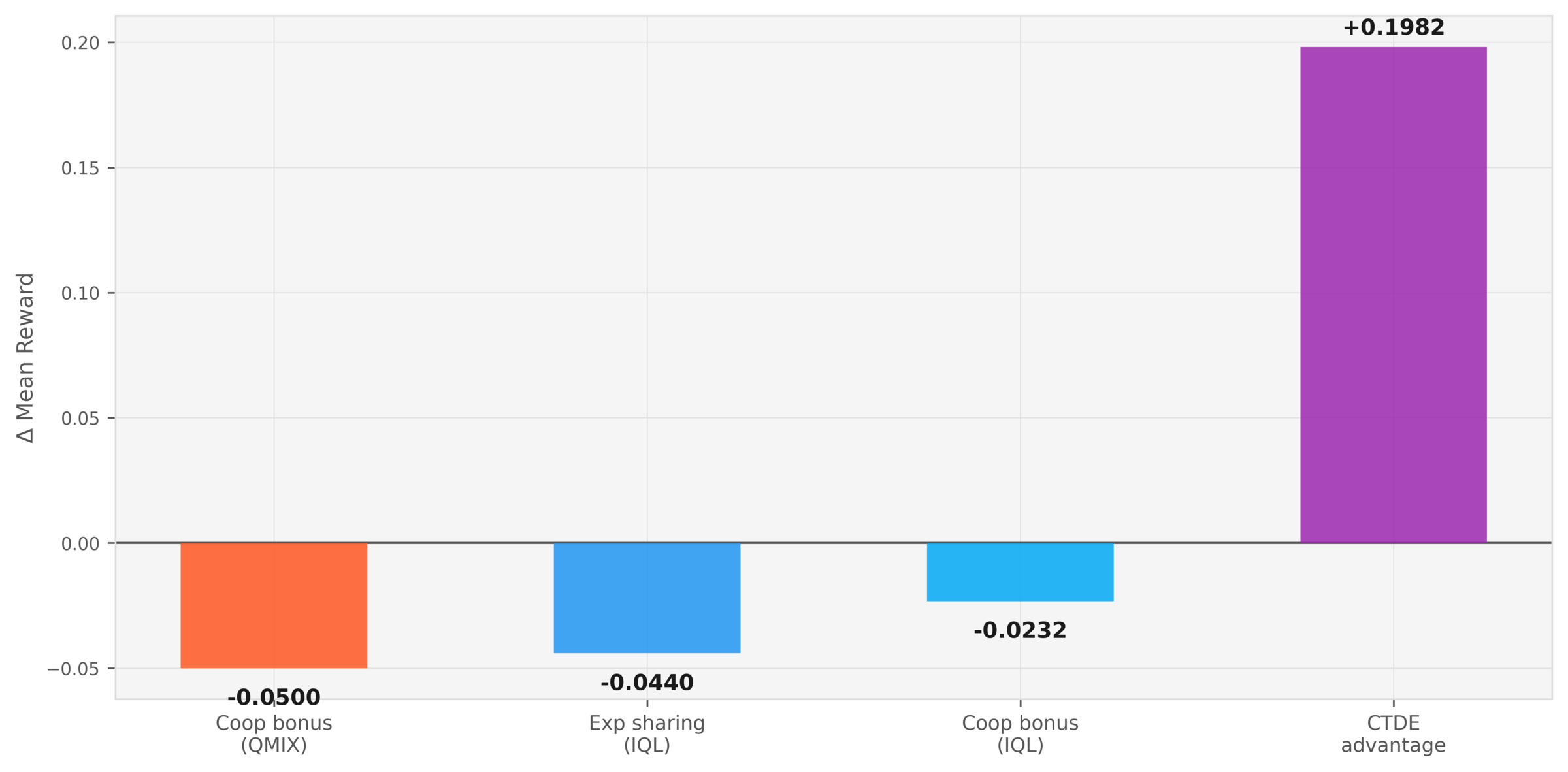}
\caption{Component effects}
\end{subfigure}
\caption{(\textbf{a}) Mean reward and accept rate for the five ablation variants (seed~=~42, 50 episodes) and (\textbf{b}) estimated contribution of each cooperative component to mean reward ($\Delta = \text{variant\_with} - \text{variant\_without}$). Positive values denote an increase in reward after adding the component; negative values denote a decrease. The only positive component effect in this ablation is associated with CTDE ($+0.198$, A0 vs.\ A3, the single-factor contrast).\label{fig:ablation}}
\end{figure}

\begin{figure}[!htb]
\centering
\includegraphics[width=\textwidth]{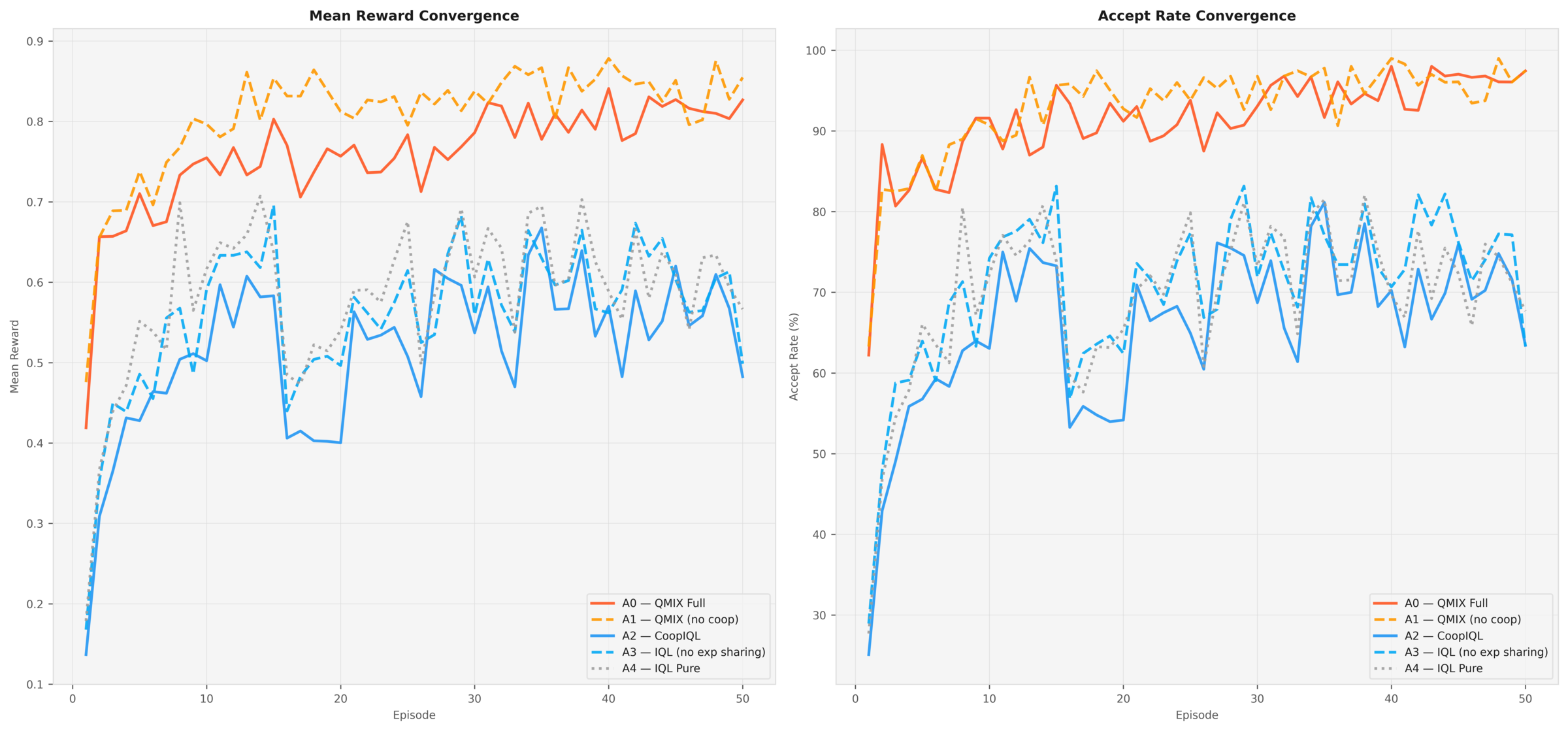}
\caption{Training trajectories of the five ablation variants (seed~=~42). The QMIX configurations (A0, A1) reach stable reward levels early in training and remain above the IQL configurations, which stay volatile throughout the 50 episodes without a clear ordering in variance among A2--A4.\label{fig:contributions}}
\end{figure}

The ablation in Table~\ref{tab:ablation} and Figure~\ref{fig:ablation} reads the performance gap as controlled contrasts rather than a single aggregate difference. The cleanest single-factor contrast is A0 vs.\ A3, which holds $L_{\text{coop}}$ and $R_{\text{exp}}$ fixed and changes only the algorithm: CTDE raises mean reward by $\Delta = +0.198$ ($d = +2.46$, Figure~\ref{fig:contributions}), considerably more than the auxiliary cooperative mechanisms (the direct A0 vs.\ A2 contrast gives a similar $\Delta = +0.242$, but also changes $R_{\text{exp}}$).

A0 vs.\ A1 shows the opposite pattern: removing the cooperative bonus raises QMIX reward by $0.050$ ($d = 0.729$, from $+0.759$ to $+0.809$), suggesting the explicit load-balance term is redundant once the monotonic mixer is trained on the joint reward signal, adding a second, partially overlapping objective to a factorization already able to encode station-load dependencies.

The IQL variants point the same way: suppressing experience sharing ($R_{\text{exp}} = 0$) raises IQL reward from $+0.517$ (A2) to $+0.561$ (A3, $\Delta = +0.044$, $d = 0.473$), while A3 vs.\ A4 shows little additional change from the cooperative bonus alone ($d = 0.252$). The measured cooperative gain is thus attributable to the CTDE architecture, not to the two heuristic cooperation terms.

\section{Discussion}
\label{sec:discussion}

\subsection{Coordination Failure Under Partial Observability}

The $21.6$\,pp acceptance-rate gap in Table~\ref{tab:main} is best read through partial observability: an independent station can make a decision that is reasonable with respect to its own queue and occupancy yet suboptimal for the district. A station near saturation may reject a vehicle although unused capacity exists elsewhere, since that information is outside its local observation and no reassignment channel exists, so the local policy cannot exploit complementarities between stations. Meanwhile each station's policy is updated while the others keep changing, exposing learning to a moving set of transition conditions --- precisely the situation the MARL literature identifies as difficult for decentralized learning under partial information~\citep{ref-hernandez2019}.

This is consistent with the convergence behavior of Figure~\ref{fig:convergence}: under \emph{High} flux, IQL does not merely converge slowly, its trajectory stays irregular even after many episodes, suggesting interference among policies trained separately on coupled outcomes. QMIX reduces this because station-level updates are evaluated through a mixer conditioned on global state, indicating that charging assignment is closer to a coupled Dec-POMDP than to independent station-control tasks~\citep{ref-oliehoek,ref-rashid2018}.

\subsection{Why the Cooperative Bonus Can Reduce Performance}

The A0--A1 contrast separates two possible sources of coordination: the learned mixer and the explicit load-balance term. The term was introduced to guide agents toward better network utilization, yet removing it improves the QMIX variant --- not because load information is irrelevant, but because it is already available to the mixer through the global state vector $\mathbf{g}$. In the independent-learner case, Equation~(\ref{eq:coop}) supplies information otherwise absent from each station's local update; for QMIX, the mixer already receives the network load state during training, so load balancing is represented through the joint TD objective itself, and a separate cooperative reward adds an optimization pressure that is close to, but not identical with, the learned value factorization --- consistent with the performance reduction observed for A0. For work combining MARL with heuristic shaping or forecasting modules~\citep{ref-korkas2024,ref-mishra2025,ref-wang2025}, this is a reminder that each added cooperative term should be validated in the presence of the chosen architecture, not assumed beneficial.

\subsection{The Effect of Experience Sharing}

The experience-sharing result reflects a mismatch between data volume and data relevance: sharing transitions increases the information available to each independent learner, but the extra samples come from stations whose operating conditions differ (fast, medium, and slow chargers differ in service duration, queue formation, and demand exposure), so a transition from a fast charger need not be a useful update for a medium-speed one. When $R_{\text{exp}} = 0.15$, each Q-table is partially updated from observations outside its own regime, mixing several station-specific dynamics into a policy that may no longer fit any one station's conditions --- explaining why removing experience sharing improves IQL reward in the A2--A3 contrast: experience transfer moves data between agents without necessarily transferring a useful decision structure.

\subsection{Coordination Effects Across Heterogeneous Stations}

Figure~\ref{fig:stations} shows the improvement is not uniform across station types: under IQL, acceptance rate spans $11.5$\,pp (from $65.7\%$ at Med-1 to $77.2\%$ at Slow-1), reflecting workload differences from station capacity and service duration. Medium-speed stations illustrate the difficulty --- attractive relative to slow chargers but slower than fast ones to clear demand, so their queues persist and a local learner reasonably becomes cautious, which can still reduce system throughput as rejected requests displace toward other congested stations. This complements work on monitoring, infrastructure heterogeneity, and load forecasting~\citep{ref-martins2025,ref-mohsenimanesh2024}, showing heterogeneity is also a learning and coordination issue. QMIX largely removes this asymmetry: acceptance rates stay between $93.5\%$ and $94.7\%$ for all six stations at a load-balance CV close to that of IQL, so the algorithmic difference is not primarily a different balance pattern but the joint value representation's ability to serve more vehicles at a similar level of balance.

\subsection{Limitations and Future Directions}

Three limitations should be noted. The evaluation uses one urban topology and binary accept/reject actions; continuous decisions, priority classes, and multi-district networks are natural extensions. The ablation uses a single seed (42) and 50 episodes; convergence curves are already separated by episode~50, but a multi-seed ablation would strengthen the attribution claims. The QMIX mixer's complexity grows with the number of agents; future work could explore more expressive factorizations, such as the duplex dueling architecture of QPLEX~\citep{ref-wang2020qplex}, for very large deployments. Future work should examine online CTDE adaptation under time-varying demand and tariff-driven non-stationarity, and integrate assignment-level coordination with energy-side formulations such as profit optimization, V2G-aware modeling, and load-forecasting-informed control~\citep{ref-jiang2025,ref-r_ucker2022,ref-mohsenimanesh2024}, so user-facing and operator-facing objectives can be optimized jointly.

\section{Conclusions}
\label{sec:conclusions}

Across five seeds and 200 episodes, QMIX reaches a $94.0\%$ acceptance rate versus $72.4\%$ for cooperative independent \textsc{Q-Learning} ($\Delta = 21.6$\,pp), with the centralized mixing network providing coordination that cooperative reward shaping and experience sharing alone do not reproduce. In the single-seed ablation, QMIX outperforms IQL under matched cooperative terms ($\Delta = +0.198$, $d = 2.46$, A0 vs.\ A3); this descriptive contrast suggests, but does not isolate, a CTDE contribution, and multi-seed ablations would be needed to confirm it.

From an operational perspective, the findings support the adoption of CTDE-based coordination in multi-station EV charging systems, rather than exclusive reliance on heuristic cooperation added to independent learners. From a methodological perspective, GradSign --- the fraction of training steps at which each agent's Q-value gradient sign agrees with the mixed Q-value's gradient sign --- equals $1.000$ over 200 episodes in this binary-action domain, consistent with, though not a formal proof of, the IGM condition.

\noindent\textbf{Funding:} This research received no external funding. \textbf{Data Availability Statement:} Data available on request from the authors. \textbf{Conflicts of Interest:} The authors declare no conflict of interest.

\bibliographystyle{unsrtnat}
\bibliography{ICATH2026_paper}

\end{document}